\documentclass[13pt,a4paper,epsf,bezier,final]{amsart}
\usepackage[margin=1.9in]{geometry}
\usepackage{amsmath,amsthm,amsfonts,amssymb,txfonts,graphicx,verbatim,siunitx}
\usepackage{bbm}
\usepackage{slashed}
\usepackage[backref]{hyperref}
\hypersetup{
colorlinks=true,
linkcolor=red,
citecolor=blue,
urlcolor=red
}
\usepackage{color}
\usepackage{tikz}

\theoremstyle{plain}

\numberwithin{sblmm}{thrm}
\numberwithin{equation}{section}

\renewcommand{\phi}{\varphi}

\def\Q.E.D{\hfill$\square$}

\newtheorem{theorem}{Theorem}
\newtheorem{lemma}{Lemma}
\newtheorem{corollary}{Corollary}
\newtheorem{definition}{Definition}

\newtheorem{remark}[theorem]{Remark}

\begin{document}
\title{On Baker-Gill-Solovay Oracle Turing Machines and Relativization Barrier}
\author{Tianrong Lin}
\address{National Hakka University\footnote{Still in the imagination of establishment.},  China}

\begin{abstract}
This work analyses the so-called ``Relativization Barrier" with respect to the Baker-Gill-Solovay oracle Turing machine. We show that the {\em diagonalization} technique is a valid mathematical proof technique, but it has some prerequisites when referring to the ``relativization barrier".

\end{abstract}
\maketitle
\tableofcontents
\section{Introduction}\label{sec_introduction}
   For any oracle $X$, we denote by $\mathcal{P}^X$ the class of languages recognized by polynomial-time deterministic Turing machines with oracle $X$, and we denote by $\mathcal{NP}^X$ the class of languages accepted by polynomial-time nondeterministic Turing machines with oracle $X$. Similarly, let $P^X$ denote the set of all polynomial-time deterministic Turing machines with oracle $X$, and $NP^X$ the set of all polynomial-time nondeterministic Turing machines with oracle $X$, respectively.

    In 1975, Baker, Gill, and Solovay \cite{BGS75} presented a proof of that: \footnote{ The proof is via $\mathcal{PSPACE}=\mathcal{NPSPACE}$, whose proof is by the fact that the space is re-useable but not valid for time, obviously.}
    \begin{eqnarray*}
      \mbox{There is an oracle $A$ for which } \mathcal{P}^A=\mathcal{NP}^A.
    \end{eqnarray*}

    Almost all complexity theory experts notice that the proof technique used to prove  $\mathcal{P}\neq EXP$, i.e., {\it diagonalization}, would also apply verbatim if we added an arbitrary oracle $O$. Thus, for any oracle $O$, we have $\mathcal{P}^O\neq EXP^O$. However, if we used similar techniques to show that $\mathcal{P}\neq\mathcal{NP}$, then it would also follow that $\mathcal{P}^O\neq \mathcal{NP}^O$ for all oracle $O$, which contradicts the above result. This is the so-called ``Relativization Barrier," which most complexity theory experts think that any proof technique leading to $\mathcal{P}\neq\mathcal{NP}$ should overcome. This viewpoint is somewhat reasonable but not absolutely correct. Personally, the motivation of \cite{BGS75} is to prove $\mathcal{P}\neq\mathcal{NP}$ via relativize, i.e., a proof technique invariant to adding oracles. Indeed, if for all oracles $O$ we can prove that $\mathcal{P}^O\neq\mathcal{NP}^O$, then it is clear that $\mathcal{P}\neq\mathcal{NP}$. But, the inverse is not necessary true, i.e., $\mathcal{P}^O\neq\mathcal{NP}^O$ for all oracle $O$ is not a necessary and sufficient condition for $\mathcal{P}\neq\mathcal{NP}$. \footnote{ For example, in the case where the set of $P^O$ is not enumerable, as will be explained below in detail.}

    Now let us return to the proof technique of {\it diagonalization} again. Cantor's diagonal process, also called the diagonalization argument, was published in 1891 by Georg Cantor \cite{Can91} as a mathematical proof that there are infinite sets which cannot be put into one-to-one correspondence with the infinite set of positive numbers, i.e., $\mathbb{N}_1$ defined in the following context. The technique of diagonalization was first used in computability theory in the 1930's by Turing \cite{Tur37} to show that there exists undecidable language. In Computational Complexity, in the seminal paper \cite{HS65}, Hartmanis et al. adapted Turing's diagonalization proof to give time hierarchy. For more summarization, please consult \cite{For00}. On the other hand, some experts or textbooks \cite{AB09} in computational complexity think that ``diagonalization" is any technique that relies solely upon the following properties of Turing machines:

    \begin{description}
      \item[I.] {The existence of an effective representation of Turing machines by strings.}
      \item[II.] {The ability of one Turing machine simulate any other without much overhead in running time or space.}
    \end{description}

    Thus, Arora et al. \cite{AB09} think that these properties are also applicable for oracle Turing machines, and further think that to show $\mathcal{P}\neq\mathcal{NP}$ requires some other properties in addition to I and II. In fact, this kind of viewpoint is not absolutely correct either. We will demonstrate the points when using diagonalization such that we can overcome the so-called ``Relativization Barrier."

    In addition to the ``Relativization Barrier," some experts in relativized worlds also developed the so-called Algebrization barrier \cite{AW09}. A more detailed discussion of this is in \cite{AB18}.

    For convenience, let us fix the notation $\mathbb{N}_1=\{1,2,3,\cdots\}$ where $+\infty\not\in\mathbb{N}_1$ and define the notion of an enumeration as follows:

    \begin{definition}[\cite{Rud76}, p. 27, Definition 2.7]\label{definition1}\footnote{ In Cantor's terminology, the enumeration of something is the ``sequence" of something. We should be clear that only enumerable sets have enumerations. And by the term ``{\it enumerable}", Turing refers to \cite{Hob21}, p. 78. That is, {\it an aggregate (i.e., set) which contains an indefinitely great number of elements is said to be enumerable, or countable, when the aggregate is such that a $(1,1)$ correspondence can be established between the elements and the set of integral numbers $1$, $2$, $3$, $\cdots$}, i.e., $\mathbb{N}_1$. We can simply deem that an enumeration of an enumerable set $T$ is just a function $e:\mathbb{N}_1\rightarrow T$ which is surjective; or equivalently, is an injective function $e':T\rightarrow \mathbb{N}_1$, meaning that every element in $T$ corresponds to a different element in $\mathbb{N}_1$. See \cite{Tur37}, Section of {\it Enumeration of computable sequences}.}
       By an enumeration of set $T$, we mean a function $e$ defined on the set $\mathbb{N}_1$ of all positive integers. If $e(n)=x_n\in T$, for $n\in\mathbb{N}_1$, it is customary to denote the enumeration $e$ by the symbol $\{x_n\}$, or sometimes by $x_1$, $x_2$, $x_3$, $\cdots$. The values of $e$, that is, the elements $x_n\in T$, are called the {\it terms} of the enumeration.
    \end{definition}

    Our main result about oracle Turing machines is {\bf to convince the experts} who argue that the ``Relativization Barrier" is a real barrier that should overcome when proving $\mathcal{P}\neq\mathcal{NP}$, we show the following important theorem about oracle Turing machines, which is to demonstrate that the ``Relativization Barrier" is not really a barrier. To do this, of course, we first suppose without loss of generality that polynomial-time deterministic (nondeterministic) Turing machines with oracles can be effectively represented as strings (i.e., the above property {\bf I}), and further, there are universal nondeterministic Turing machines with oracles that can simulate and flip the answers of other deterministic Turing machines with oracles without much overhead in running time or space (i.e., the above property {\bf II} \footnote{ This property should be divided into two sub-properties, see assumptions (II) and (III) of footnote $5$.}).

    \begin{theorem}\label{theorem1}\footnote{
      The proof of this theorem, in fact, is similar to the proof of Cantor's theorem: there are infinite sets that can not be put into one-to-one correspondence with the set of positive integers, i.e., $\mathbb{N}_1$. See \cite{Gra94}. Furthermore, the argument of this theorem lies in the assumptions that: (I) polynomial-time deterministic (nondeterministic) Turing machines with oracle can be effectively represented as strings; (II) a universal nondeterministic Turing machine with oracle exists that can simulate and flip the answers of other deterministic Turing machines with oracle; and (III) the simulation of a universal nondeterministic Turing machine with oracle to any deterministic Turing machine with oracle can be done within $O(T(n)\log T(n))$ moves, where $T(n)$ is the time complexity of the simulated deterministic Turing machine with oracle.
    }
     Let $P^O$ be the set of all polynomial-time deterministic Turing machines with oracle $O$. If $\mathcal{P}^O=\mathcal{NP}^O$, then the set $P^O$ is not enumerable. That is, the cardinality of $P^O$ is larger than that of $\mathbb{N}_1$ ( card $P^O>$card $\mathbb{N}_1$).
   \end{theorem}

   It follows from the above Theorem \ref{theorem2} that

   \begin{corollary}\label{corollary1}\footnote{ In Turing's way, he first assumes that the computable sequences are enumerable, then applies the diagonal process. See \cite{Tur37}, Section $8$ of {\it Application of the diagonal process}.}
    If a set $T$ (of oracle Turing machines) is enumerable, then the {\it diagonalization} technique is applicable.
   \end{corollary}

   Finally, we show an additional result, i.e., we will prove that the following Theorem \ref{theorem2}:

   \begin{theorem}\label{theorem2}
     For any time-constructible function $T(n)\geq n$, there exists a language $L_d^T$ that is accepted by a universal nondeterministic Turing machine of time complexity $c\dot T(n)$ for some constant $c>0$ but not by any deterministic Turing machines of time complexity $O(T(n)$. To put it another way, $DTIME [T(n)] \subset NTIME [T(n)]$.
   \end{theorem}

   We emphasize that we are considering only the growth rates of time functions. That is, for our purposes, the time functions $T(n)$ and $cT(n)$ are the same, for any constant $c>0$.

   Let $T(n)=n^i$ for any fixed $i\in\mathbb{N}_1$, then it is clear that $n^i$ is time-constructible. Hence, from the above Theorem \ref{theorem2} we can obtain the following consequence:

   \begin{corollary}\label{corollary2}
      For any fixed $k\in\mathbb{N}_1$, there exists a language $L_d^k$ that is accepted by an universal nondeterministic Turing machine of time complexity $O(n^k)$ but not by any deterministic Turing machines of time complexity $O(n^k)$. That is, $DTIME[n^k]\subset NTIME[n^k]$.
   \end{corollary}

    The rest of this work is organized as follows: For the convenience of the reader, in the next Section we review some notions closely associated with our discussions and fix some notation we will use in the following context. In Section \ref{sec_proofofmaintheorem}, we present a proof of our main result. In Section \ref{sec_an_additional_result}, we give the proof of Theorem \ref{theorem2}. Finally, we draw some conclusions in the last section.

   \section{Preliminaries}\label{sec_preliminaries}

    In this Section, we describe the notation and notions needed in the following context.

    The computation model we use in this Section is the {\it query machines}, or oracle machines, which is an extension of the multitape Turing machine as described in \cite{AHU74}, i.e., Turing machines that are given access to a black box or ``oracle" that can magically solve the decision problem for some language $O\subseteq\{0,1\}^*$. The machine has a special {\it oracle tape} on which it can write a string $q\in\{0,1\}^*$ and in one step gets an answer to a query of the form ``Is $q$ in $O$?", which can be repeated arbitrarily often with different queries. If $O$ is a difficult language (say, one that cannot be decided in polynomial time, or is even undecidable), then this oracle gives the Turing machine additional power. We quote its formal definition as follows:

    \begin{definition}(\cite{AB09}, Oracle Turing machines)\label{definition2}
       An {\it oracle} Turing machine is a Turing machine $M$ that has a special read-write tape we call $M$'s {\it oracle tape} and three special states $q_{query}$, $q_{yes}$, $q_{no}$. To execute $M$, we specify in addition to the input a language $O\subseteq\{0,1\}^*$ that is used as the {\it oracle} for $M$. Whenever during the execution $M$ enters the state $q_{query}$, the machine moves into the state $q_{yes}$ if $q\in O$ and $q_{no}$ if $q\not\in O$, where $q$ denotes the contents of the special oracle tape. Note that, regardless of the choice of $O$, a membership query to $O$ counts only as single computation step. If $M$ is an oracle machine, $O\subseteq\{0,1\}^*$ a language, and $x\in\{0,1\}^*$, then we denote the output of $M$ on input $x$ and with oracle $O$ by $M^O(x)$.
    \end{definition}

    The above Definition \ref{definition2} is for {\it Deterministic Oracle Turing Machines}. The {\it Nondeterministic Oracle Turing machine} can be defined similarly.

    If for every input $w$ of length $n$, all computations of $M^X$ end in less than or equal to $T(n)$ moves, then $M^X$ is said to be a $T(n)$ {\it time-bounded (nondeterministic) Turing machine with oracle $X$}, or of time complexity $T(n)$. The family of languages of deterministic time complexity $T(n)$ is denoted by DTIME$^X$[$T(n)$]; the family of languages of nondeterministic time complexity $T(n)$ is denoted by NTIME$^X$[$T(n)$]. The notation $\mathcal{P}^X$ and $\mathcal{NP}^X$ are defined to be the class of languages:
    \begin{eqnarray*}
      \mathcal{P}^X=\bigcup_{k\in \mathbb{N}_1}DTIME^X[n^k]
    \end{eqnarray*}
    and
    \begin{eqnarray*}
     \mathcal{NP}^X=\bigcup_{k\in\mathbb{N}_1}NTIME^X[n^k].
    \end{eqnarray*}

   \section{``Relativization Barrier" is not a real barrier, for $P^O$ is not enumerable if $\mathcal{P}^O=\mathcal{NP}^O$}\label{sec_proofofmaintheorem}

   We now emphasizing that {\bf (I)} the polynomial-time deterministic (nondeterministic) Turing machines with oracle can be effectively represented as strings, and further suppose that {\bf (II)} there are universal nondeterministic Turing machine with oracle that can simulate any other without much overhead in running time and flip answer of other deterministic Turing machines with oracle (See footnote $6$), which satisfies {\bf I} and {\bf II} mentioned in Section \ref{sec_introduction}.We proceed to prove Theorem \ref{theorem1}:

   \subsection{Proof of Theorem \ref{theorem1}}

   \vskip 0.3 cm
 We show Theorem \ref{theorem1} by contradiction. Suppose to the contrary that the set $P^O$ is enumerable, or in other words, the cardinality of $P^O$ is less than or equal to that of $\mathbb{N}_1$. Then we have an enumeration $e:\mathbb{N}_1\rightarrow P^O$. Next, we construct a five-tape universal nondeterministic Turing machine $M_0^O$ with oracle $O$ which operates as follows on an input string $x$ of length of $n$:
   \begin{enumerate}
       \item{ $M^O_0$ decodes the tuple encoded by $x$. If $x$ is not the encoding of some polynomial-time DTM $D^O_j$ with oracle $O$ for some $j$ then GOTO $5$, else determines $t$, the number of tape symbols used by $D^O_j$; $s$, its number of states; and $k$, its order of polynomial.\footnote{ Assume that the order of polynomial of Oracle Turing machine $D\in P^O$ is encoded into $D$.} The third tape of $D^O_0$ can be used as ``scratch" memory to calculate $t$.}
       \item{  Then $D^O_0$ lays off on its second tape $|x|$ blocks of $\lceil\log t\rceil$ cells each, the blocks being separated by single cell holding a marker $\#$, i.e., there are $(1+\lceil\log t\rceil)n$ cells in all where $n=|x|$. Each tape symbol occurring in a cell of $D^O_j$'s tape will be encoded as a binary number in the corresponding block of the second tape of $M^O_0$. Initially, $M^O_0$ places $D^O_j$'s input, in binary coded form, in the blocks of tape $2$, filling the unused blocks with the code for the blank.}

       \item{ On tape $3$, $M^O_0$ sets up a block of $\lceil(k+1)\log n\rceil$ cells, initialized to all $0$'s. Tape $3$ is used as a counter to count up to $n^{k+1}$.}

       \item{ On tape $4$, $M^O_0$ reads and writes the contents of oracle tape of $D^O_j$. That is, tape $4$ is the oracle tape of $M^O_0$.}

       \item{ By using nondeterminism, $M^O_0$ simulates $D^O_j$, using tape $1$, its input tape, to determine the moves of $D^O_j$ and using tape $2$ to simulate the tape of $D^O_j$, further using tape $4$ to simulate the oracle tape of $D^O_j$. The moves of $D^O_j$ are counted in binary in the block of tape $3$, and tape $5$ is used to hold the state of $D^O_j$. If $D^O_j$ accepts, then $M^O_0$ halts without accepting. $M^O_0$ accepts if $D^O_j$ halts without accepting, or if the counter on tape $3$ overflows, $M^O_0$ halts without accepting.}
       \item{ Since $x$ is not encoding of some single-tape DTM with oracle $O$. Then $M^O_0$ sets up a block of $\lceil 2\times\log n\rceil$ cells on tape $3$, initialized to all $0$'s. Tape $3$ is used as a counter to count up to $n^2$. By using its nondeterministic choices, $M^O_0$ moves as per the path given by $x$. The moves of $M^O_0$ are counted in binary in the block of tape $3$. If the counter on tape $3$ overflows, then $M^O_0$ halts. $M^O_0$ accepts $x$ if and only if there is a computation path from the start state of $M^O_0$ leading to the accept state and the total number of moves can not exceed $n^2$, so is within $O(n)$. Note that the number of $2$ in $\lceil 2\times\log n\rceil$ is fixed, i.e., it is default.}
     \end{enumerate}

     The NTM $M^O_0$ described above with oracle $O$ is of time complexity $S(n)=O(n^m)$ for any $m\in\mathbb{N}_1$, and it, of course, accepts some language $L^O_d$.

     Suppose now $L^O_d$ were accepted by $i$-th DTM $D^O_i$ with oracle $O$ in the enumeration $e$ which is of time complexity $T(n)=O(n^k)$. Let $D^O_i$ have $s$ states and $t$ tape symbols. Since \footnote{ See (III) of footnote $4$.}
     \begin{eqnarray*}
       \lim_{n\rightarrow\infty}&&\frac{T(n)\log T(n)}{n^{k+1}}\\
       &=&\lim_{n\rightarrow\infty}\frac{cn^k(\log c+k\log n)}{n^{k+1}}\quad\mbox{( for some constant $c>0$ )}\\
       &=&\lim_{n\rightarrow\infty}\Big(\frac{c\log c n^k}{n^{k+1}}+\frac{ckn^k\log n}{n^{k+1}}\Big)\\
       &=&0\\
       &<&1.
     \end{eqnarray*}

     So, there exists a $N_0>0$ such that for any $N\geq N_0$,
     \begin{eqnarray*}
        T(N)\log T(N)&<&N^{k+1}
     \end{eqnarray*}

     which implies that for a sufficiently long $w$, say $|w|\geq N_0$, and $D^O_w$ denoted by such $w$ is $D^O_i$, we have that

     \begin{eqnarray*}
        T(|w|)\log T(|w|)&<&|w|^{k+1}.
     \end{eqnarray*}

     Thus, on input $w$, $M^O_0$ has sufficient time to simulate $D^O_w$ and accepts if and only if $D^O_w$ rejects (In simulation of a polynomial-time DTM with oracle $O$, $M^O_0$ only turns off mandatorily when the counter on tape $3$ overflows, i.e., the counter $\geq N^{k+1}$). But we assumed that $D^O_i$ accepted $L^O_d$, i.e., $D^O_i$ agreed with $M^O_0$ on all inputs. We thus conclude that $D^O_i$ does not exist, i.e., $L^O_d$ not accepted by any machine in the enumeration $e$, in other words, $L^O_d\not\in \mathcal{P}^O$.

     Next we show that $L^O_d\in \mathcal{NP}^O$. Define the family of languages $\{L^O_{d,i}\}_{i\in\mathbb{N}_1}$ as follows:
     \begin{eqnarray*}
         L^O_{d,i}&\triangleq&\mbox{ language accepted by $M^O_0$ running within time $O(n^i)$ for fixed $i$ where $i\in\mathbb{N}_1$.}\\
         &&\mbox{ That is, $M^O_0$ turns off mandatorily when its moves made during the computation}\\
         &&\mbox{ exceeds $n^{i+1}$.}
     \end{eqnarray*}
     Then by construction, since $M^O_0$ runs within time $O(n^i)$ for any fixed $i\in\mathbb{N}_1$, we thus have that
     \begin{eqnarray}\label{eq7}
        L^O_d&=&\bigcup_{i\in\mathbb{N}_1}L^O_{d,i}.
     \end{eqnarray}
     Furthermore,
     \begin{eqnarray*}
       L^O_{d,i}\subseteq L^O_{d,i+1}\quad\mbox{ for each fixed $i\in\mathbb{N}_1$}
     \end{eqnarray*}
     since for any word $w\in L^O_{d,i}$ accepted by $M^O_0$ within $O(n^i)$ moves, it surely can be accepted by $M^O_0$ within $O(n^{i+1})$ moves, i.e., $w\in L^O_{d,i+1}$. This gives that for any fixed $i\in\mathbb{N}_1$,
     \begin{eqnarray}\label{eq8}
       L^O_{d,1}\subseteq L^O_{d,2}\subseteq\cdots\subseteq L^O_{d,i}\subseteq L^O_{d,i+1}\subseteq\cdots
     \end{eqnarray}
     Note further that for any fixed $i\in\mathbb{N}_1$, $L^O_{d,i}$ is accepted by a nondeterministic Turing machine $M^O_0$ with oracle $O$ within $O(n^i)$ moves, we thus obtain that
     \begin{eqnarray}\label{eq9}
       L^O_{d,i}\in NTIME^O[n^i]\subseteq\mathcal{NP}^O\quad\mbox{ for any fixed $i\in\mathbb{N}_1$}.
     \end{eqnarray}

     From (\ref{eq7}), (\ref{eq8}) and (\ref{eq9}) we deduce that
      \begin{eqnarray*}
       L^O_d\in \mathcal{NP}^O.
      \end{eqnarray*}

     To summarize, we obtain that
     \begin{eqnarray*}
         \mathcal{P}^O\neq\mathcal{NP}^O,
     \end{eqnarray*}
     which contradicts the assumption that $\mathcal{P}^O=\mathcal{NP}^O$. So, we can conclude that the set of $P^O$ is not enumerable.\footnote{ In other words, for any function $e$ : $\mathbb{N}_1\rightarrow P^O$, there always exists an element $M^O\in P^O$, such that, for any $i\in\mathbb{N}_1$, $e(i)\neq M^O$.} This completes the proof of Theorem \ref{theorem2}. \Q.E.D

     \vskip 0.3 cm
     \begin{remark}\label{remark6}
        In fact, under the condition that $\mathcal{P}^O=\mathcal{NP}^O$, we can assume that the set $P^O$ of all polynomial-time deterministic Turing machines with oracle $O$ is enumerable. Then we can show that for any enumeration $e:\mathbb{N}_1\rightarrow P^O$, there is always a machine $D^O_S$ in $P^O$ such that $e(i)\neq D^O_S$ for any $i\in\mathbb{N}_1$ which contradicts the assumption that $P^O$ is enumerable.\footnote{ The language accepted by machine $D^O_S$ differs from the languages accepted by all of the polynomial-time deterministic Turing machines with oracle $O$ in the enumeration, but it lies in $P^O$ since $\mathcal{P}^O=\mathcal{NP}^O$.}
        As a result, we are unable to diagonalize over the set of $P^O$, as Cantor \cite{Can91} was unable to put all real numbers in the open interval $(0,1)$ into the slots indexed by all $i\in\mathbb{N}_1$.\footnote{ For a more detailed comparison, the reader could consult the second proof (due to Cantor) that the continuum is not enumerable. See \cite{Hob21}, p. 82.}
     \end{remark}
       \vskip 0.3 cm

      We can conclude that diagonalization by a universal nondeterministic Turing machine with oracle $O$ is not suitable for separating $\mathcal{P}^O$ and $\mathcal{NP}^O$ if $\mathcal{P}^O=\mathcal{NP}^O$, because in this case, we can always construct a machine $D^O_S$ so that, for any function $e:\mathbb{N}_1\rightarrow P^O$, there is no $i\in\mathbb{N}_1$ such that $e(i)=D^O_S$. In other words, the set of $P^O$ is not enumerable. Equivalently, the cardinality of $P^O$ is greater than that of $\mathbb{N}_1$. This is the most significant difference. In a nutshell, the fact that the set of $P$ is enumerable is an important prerequisite for the application of diagonalization.

   \section{An additional result}\label{sec_an_additional_result}

   This similar technique, i.e., using nondeterministic Turing machine to diagonalize against deterministic Turing machines, is used for the first time in the author's work \cite{Lin21} to separate two different complexity classes $DSPACE [S(n)]$ and $NSPACE [S(n)]$ for some space-constructible function $S(n))\geq\log n$. The inspirations are drawn from two facts: (1) the author was reading the proof of the space hierarchy for deterministic Turing machines, i.e., Theorem 11.1 in \cite{AHU74}; and (2) the author was considering how to resolve a longstanding open question in {\em automata theory}, i.e., the {\em LBA question}. Then the thought occurred to me to use a universal nondeterministic Turing machine to diagonalize a list of deterministic Turing machines of space complexity, say $S(n)\geq\log n$.

   To prove the corollary \ref{corollary2} or the more general Theorem \ref{theorem2}, we need the following Lemma \ref{lemma1}. However, there is lack of the formal proof of the following Lemma \ref{lemma1} due to the capability of the author.

   \begin{lemma}(\cite{AB09}, p. 64, Exercises 2.6 (b))\label{lemma1}\footnote{ We just think this lemma is correct. If it is incorrect, then Theorem \ref{theorem2} is invalid.}
      An universal nondeterministic Turing machine $U_N$ can simulate a nondeterministic Turing machine $M_N$ in time $C\cdot T(n)$ where $M_N$ runs within time $T(n)$ and $C$ is a constant depending only on the machine $M_N$.
   \end{lemma}

   Since a deterministic Turing machine is also a nondeterministic Turing machine, by definition. As a result of Lemma \ref{lemma1}, we believe that a universal nondeterministic Turing machine can simulate a deterministic Turing machine in time $O(T(n))$, where the deterministic Turing machine runs within time $O(T(n))$.

    We can now design a four-tape NTM $M^0$ which treats its input string $x$ both as an encoding of a deterministic Turing machine $M$ and also the input to $M$. We shall have $M^0$ determine whether the deterministic Turing machine $\widehat{M}_i$ \footnote{ See \cite{Lin21$^+$} how to encode a deterministic Turing machine into a binary string.} of time complexity $T(n)$ accepts the input $x$ without using more than time $T(n)$. If $\widehat{M}_i$ accepts $x$ within in time $T(n)$, then $M_0$ does not. Otherwise, $M^0$ accepts $x$. Thus, for all $i$, $M^0$ disagrees with the behavior of $\widehat{M}_i$ of time complexity $T(n)$ in the $i$th of enumeration on that input $x$, i.e., the following:

    \subsection{Proof of Theorem \ref{theorem2}}
     Let $M^0$ be a four-tape NTM which operates as follows on an input string $x$ of length of $n$.
     \begin{enumerate}
       \item{ $M^0$ decodes the tuple encoded by $x$. If $x$ is not the encoding of some single-tape polynomial-time DTM $\widehat{M}_j$ for some $j$ then GOTO $5$, else determines $t$, the number of tape symbols used by $\widehat{M}_j$; $s$, its number of states. The third tape of $M^0$ can be used as ``scratch" memory to calculate $t$.}
       \item{  Then $M^0$ lays off on its second tape $|x|$ blocks of $\lceil\log t\rceil$ cells each, the blocks being separated by single cell holding a marker $\#$, i.e. there are $(1+\lceil\log t\rceil)n$ cells in all where $n=|x|$. Each tape symbol occurring in a cell of $\widehat{M}_j$'s tape will be encoded as a binary number in the corresponding block of the second tape of $M^0$. Initially, $M^0$ places $\widehat{M}_j$'s input, in binary coded form, in the blocks of tape $2$, filling the unused blocks with the code for the blank.}

       \item{ On tape $3$, $M^0$ sets up a block of $\lceil \log (C_{\widehat{M}_j})+\log T(n)\rceil$ cells, initialized to all $0$'s. Tape $3$ is used as a counter to count up to $C_{\widehat{M}_j}T(n)$ where $C_{\widehat{M}_j}$ is a constant depending only on $\widehat{M}_j$.}

       \item{ By using nondeterminism, $M^0$ simulates $\widehat{M}_j$, using tape $1$, its input tape, to determine the moves of $\widehat{M}_j$ and using tape $2$ to simulate the tape of $\widehat{M}_j$. The moves of $\widehat{M}_j$ are counted in binary in the block of tape $3$, and tape $4$ is used to hold the state of $\widehat{M}_j$. If $\widehat{M}_j$ accepts, then $M^0$ halts without accepting. $M^0$ accepts if $\widehat{M}_j$ halts without accepting. If the counter on tape $3$ overflows, $M^0$ halts mandatorily without accepting.}
       \item{ Since $x$ is not encoding of some single-tape DTM. Then $M^0$ sets up a block of $\lceil 2\times\log n\rceil$ cells on tape $3$, initialized to all $0$'s. Tape $3$ is used as a counter to count up to $n^2$. By using its nondeterministic choices, $M^0$ moves as per the path given by $x$. The moves of $M^0$ are counted in binary in the block of tape $3$. If the counter on tape $3$ overflows, then $M^0$ halts. $M^0$ accepts $x$ if and only if there is a computation path from the start state of $M^0$ leading to the accepting state and the total number of moves can not exceed $n^2$, so is within $O(n)$. Note that the number of $2$ in $\lceil 2\times\log n\rceil$ is fixed, i.e. it is default.}
     \end{enumerate}

     The NTM $M^0$ described above is of time complexity $CT(n)$ where $C$ is a constant. Because $M^0$ turns off mandatorily when the total number of moves made by $M^0$ exceeds or equal to $C_{\widehat{M}_j}T(|w|)$ for input $w$ if $w$ encodes some single-tape deterministic Turing machines of time complexity $T(n)$. It of course accepts some language $L^T_d\in NTIME[T(n)]$.

     Suppose now $L^T_d$ were accepted by some DTM $\widehat{M}_j$ in the enumeration which is of time complexity $T(n)$. Then we may assume that $\widehat{M}_j$ is a single-tape DTM. Let $\widehat{M}_j$ have $s$ states and $t$ tape symbols. Since $\widetilde{M_j}$ \footnote{ $\widetilde{M}_j$ denotes the set of binary strings which encodes $\widehat{M}_j$. We know that we may prefix $1$'s at will to find larger and larger integers representing the same set of quintuples of the same DTM $M_j$ (see \cite{Lin21$^+$}), thus there are infinitely binary strings of sufficiently long which represents DTM $M_j$.} appears infinitely often in the enumeration, and by Lemma \ref{lemma1}, the simulation can be done within $C_{\widetilde{M}_j}T(n)$ :
     \begin{eqnarray*}
       \lim_{n\rightarrow\infty}&&\frac{T(n)}{C_{\widetilde{M}_j}T(n)}\\
       &=&\frac{1}{C_{\widetilde{M}_j}}\\
       &<&1,\quad\mbox{ (of course $M^0$ is a bit slow than $\widetilde{M}_j$, so $C_{\widetilde{M}_j}>1$)}
     \end{eqnarray*}

     So, there exists a $N_0>0$ such that for any $N\geq N_0$,
     \begin{eqnarray*}
        T(N)&<&C_{\widetilde{M}_j}\cdot T(N)
     \end{eqnarray*}

     which implies that for a sufficiently long $w$, say $|w|\geq N_0$, and $M_w$ denoted by such $w$ is $\widehat{M}_j$, we have that

     \begin{eqnarray*}
        T(|w|)&<&C_{\widetilde{M}_j}T(|w|).
     \end{eqnarray*}

     Thus, on input $w$, $M^0$ has sufficient time to simulate $M_w$ and accepts if and only if $M_w$ rejects. But we assumed that $\widehat{M}_j$ accepted $L^T_d$, i.e. $\widehat{M}_j$ agreed with $M^0$ on all inputs. We thus conclude that $\widehat{M}_j$ does not exist, i.e. $L^T_d\not\in DTIME[T(n)]$. Note again that we are considering only the growth rates of time functions. That is, for our purposes, the time function $T(n)$ and $cT(n)$ are the same, for any constant $c>0$.\Q.E.D

   \section{Conclusions}\label{sec_conclusions}

    We have shown that if $\mathcal{P}^O=\mathcal{NP}^O$, then the set of $P^O$ is not enumerable. So we cannot use the diagonalization technique by a universal nondeterministic Turing machine with oracle $O$ to separate $\mathcal{P}^O$ and $\mathcal{NP}^O$. This shows that the so-called ``Relativization Barrier" is not a barrier, in fact.

\bibliographystyle{amsplain}

\vskip 0.6cm
\end{document}